\documentclass[proof]{WileyASNA-v1}

\articletype{Article Type}%

\received{xx.xx.xxxx}
\revised{xx.xx.xxxx}
\accepted{xx.xx.xxxx}

\usepackage{upgreek}

\begin{document}

\title{Effects of flares on the habitable zones of M dwarfs accessible to TESS planet detections}

\author[1]{M. Bogner*}

\author[1,2]{B. Stelzer}

\author[1]{St. Raetz}

\authormark{BOGNER \textsc{et al}}

\address[1]{\orgdiv{}, \orgname{Institut
 für Astronomie und Astrophysik
Tübingen (IAAT), Eberhard-Karls
Universität Tübingen}, \orgaddress{\state{Sand 1, D-72076 Tübingen}, \country{Germany}}}

\address[2]{\orgdiv{}, \orgname{INAF - Osservatorio Astronomico di
Palermo}, \orgaddress{\state{Piazza del Parlamento 1, I-90134,
Palermo
}, \country{Italy}}}

\corres{*\email{bogner@astro.uni-tuebingen.de}\\\\\\
{\bf This is the pre-peer reviewed version of the following article: M. Bogner, B. Stelzer, and St. Raetz (2021), Effects of flares on the habitable zones of M dwarfs accessible to TESS planet detections, which has been published in final form at doi.org/10.1002/asna.20210079. This article may be used for non-commercial purposes in accordance with Wiley Terms and Conditions for Use of Self-Archived Versions.}}

\abstract{Photometric space missions like {\it Kepler} and {\it TESS} continuously discover new exoplanets and advance the search for a second habitable world. The light curves recorded by these telescopes also reveal signs of magnetic activity, such as star spot modulation and flares, which can influence habitability. Searching for these characteristics, we analyzed {\it TESS} light curves of 112 M dwarfs selected according to the criterion that {\it TESS} can spot planet transits over their entire habitable zone (HZ). We detected 2532 flare events, occurring on 35 stars, thus the flare incidence rate is $\approx32\%$. For only $\approx11\%$ of our stars, we found rotation periods. We calculated bolometric flare energies and luminosities, flare energy frequency distributions (FFDs) and the bolometric flux reaching the HZ at the peak of the flare. We estimated the effects of flaring on the atmosphere of an Earth-like planet in the HZ in the view of both ozone depletion and the enabling of chemical reactions necessary to build ribonucleic acid (RNA). None of our targets exhibits highly energetic flares at a frequency large enough to trigger ozone depletion or RNA formation. The flux reaching the inner HZ edge during flare events goes up to $\approx 4.5$ times the quiescent solar flux at 1\,au.}


\jnlcitation{\cname{%
\author{M. Bogner}, 
\author{B. Stelzer}, and  
\author{St. Raetz}} (\cyear{2021}), 
\ctitle{title.}}


\maketitle

\section{Introduction}\label{sec:introduction}
Within the last decade, the number of confirmed exoplanets has grown from less than 700 to over 4500. This is mainly the result of the {\it Kepler} mission \citep{Borucki2010} which aimed particularly at spotting Earth-sized planets around Sun-like stars. The insights gained by this successful mission were and will be expanded by the means of several successors such as the {\it Transiting Exoplanet Survey Satellite} ({\it TESS}, \citealt{Ricker2015}), {\it CHaracterizing Exoplanet Satellite} (CHEOPS, \citealt{Benz2020}) and PLATO \citep{Rauer2016} which are designed to detect further planets or characterize the ones already known. Analogously to the {\it Kepler} and {\it K2} \citep{Howell2014} mission, {\it TESS} light curves not only reveal planet transits but also signs of stellar magnetic activity that manifest at optical wavelengths through stochastic brightness outbursts called flares and star spot modulation. {\it TESS} is a NASA satellite launched in April 2018. It completed its first nearly all-sky survey in July 2020 and continues operating for a second all-sky survey. Our work is based on 112 M dwarfs observed by {\it TESS} in 2-min. cadence during the primary mission. We analyse their flaring activity in view of the habitability of a potential planet in the habitable zone. A flaring host star is a double-edged sword regarding the development and persistence of life on such a planet: On the one hand, the increased flux during flare events can trigger chemical reactions that are necessary to build ribonucleotides which form the basis of ribonucleic acid (RNA) and therefore help the development of prebiotic chemistry (e. g. \citealt{Sutherland2015}). On the other hand, flares with energies $\geq 10^{34}$\,erg can lead to an erosion of the ozone layer of the planet's atmosphere when they occur frequently enough (e. g. \cite{Tilley2019}). As a consequence, the planet surface is no longer shielded from stellar UV radiation that can be life harming when occurring in sufficiently large doses.

We explain the details of our sample selection in Sect.~\ref{sec:sample}. Sect.~\ref{sec:lc_analysis} summarizes the most important points of our rotation period search and flare detection procedure, the results of which are given in Sect.~\ref{sec:results}. We present flare incidence rates in dependence of the spectral type and flare activity as a function of the rotation period in Sect.~\ref{subsec:flare_incidence_activity}. Sect.~\ref{subsec:flare_effects_HZ} discusses flare energy frequency distributions (FFDs) in terms of the bolometric flare energy and the flux reaching the habitable zone at the peak of each flare event. To estimate the effects of our flares on a potential planet in the habitable zone, we also calculated the abiogenesis zone for each star following \cite{Guenther2020} and compare our FFDs to the flare energies and frequencies sufficient to trigger ozone depletion according to the model analysis of \cite{Tilley2019}. Sect.~\ref{subsec:flare_toi} deals with the flare results for the {\it TESS Objects of Interest} (TOI) and confirmed planet host stars within our sample. In Sect.~\ref{sec:discussion} we summarize and discuss our results.

\section{Sample}\label{sec:sample}
We selected a subset of stars from the {\it TESS Habitable Zone Star Catalog} (HZCat, \citealt{Kaltenegger2019}) with the property that {\it TESS} can detect planet transits over the full extent of the habitable zone. The HZCat lists in total 227 stars with this property. In order to maximize the signal to noise ratio of the light curves considered in our analysis, we introduced a more restrictive magnitude cutoff than that in the HZCat and kept only stars with {\it TESS} magnitudes $T\leq11.5$\,mag. This reduces the sample to 112 M dwarfs covering a mass range from $0.17M_\odot$ to $0.70M_\odot$.

For the determination of stellar parameters, we applied the empirical relations of \cite{Mann2015}, using the coefficients from \cite{Mann2016}. Photometric data was used from the {\it Two Micron All-Sky Survey} (2MASS, \citealt{2mass}) and from {\it Gaia} DR2 \citep{GaiaDr2}. To determine $V$ band magnitudes from {\it Gaia} photometry, we used the empirical relation from \cite{Jao2018}. We derived distances from inverting {\it Gaia} DR2 parallaxes\footnote{We compared our distances obtained for our sample by inverting {\it Gaia} DR2 parallaxes to those provided by \cite{BailerJones2018}, finding a maximum difference of $\approx0.2$\,pc. This speaks well for the accuracy of the inverted parallax distances in the case of our sample.} and checked their reliability by applying the conditions suggested by \cite{Lindegren2018} in their appendix~C. For the eleven of our 112 targets that have no reliable {\it Gaia} DR2 distance according to this criterion, we used photometric distances (see Bogner et al., A\&A subm. for details). 

To obtain bolometric luminosities, $L_\text{bol,*}$, we used V band magnitudes calculated as described above, {\it Gaia} DR2 distances and bolometric corrections $BC_V$ given in the online table {\it A Modern Mean Dwarf Stellar Color and Effective Temperature Sequence} maintained by E. Mamajek\footnote{\label{note1}The table is available at\\pas.rochester.edu/\textasciitilde emamajek/EEM\_dwarf\_UBVIJHK\_colors\_Teff.txt} as an extension of \citet{Pecaut2013}. 

For further details on the collection of photometric data and the stellar parameter calculation, see Bogner et al. (A\&A subm.). 

\section{Light curve analysis}\label{sec:lc_analysis}
1276 2-minute cadence {\it TESS} light curves that we downloaded from the Mikulski Archive for Space Telescopes (MAST)\footnote{mast.stsci.edu} form the basis of our work. In the course of the analysis, we noticed that 3 of our 112 targets lie outside the aperture mask selected automatically by the pipeline developed at NASA's Science Processing Operations Center (SPOC).\footnote{The IDs of the 3 stars are TIC359676790, TIC392572237 and TIC471015740.} We therefore excluded these 3 stars from our analysis, leaving a sample of 109 targets. We used pre-search data conditioning simple aperture photometry (PDC SAP) LCs. 

\subsection{Rotation period search}\label{subsec:prot_search}
We searched for rotational modulation caused by star spots to constrain the rotation periods, $P_\text{rot}$, of our stars. To this end, we applied 3 different methods, following our previous works (\citealt{Stelzer2016}, \citealt{Raetz2020}): First, a generalized Lomb-Scargle-Periodogram (GLS) using an algorithm by \cite{Zechmeister2009}\footnote{astro.physik.uni-goettingen.de/\textasciitilde zechmeister/gls.php,
v2.3.02, released 2012-08-03}, second, an autocorrelation function (ACF) and third, a sine fit. Almost all our stars have been observed in multiple {\it TESS} sectors. For the period search, however, we treated the LC of each sector individually. The overall $P_\text{rot}$ result was obtained from the results of all sectors as their weighted mean value (see Bogner et al., A\&A submitted for details).

\subsection{Flare detection and parameters}\label{subsec:flare_detection}
Our flare detection algorithm has been described by \cite{Stelzer2016} and \cite{Raetz2020}. For details on the flare detection procedure, we refer to the descriptions given in these two references, as we provide here only a summary of the most important points. The first step of the algorithm is to create a flattened and cleaned light curve and calculate its standard deviation, $\sigma$. Then, all groups of at least 3 consecutive data points in the original LC that deviate by more than $3\sigma$ from the mean value of the flattened and cleaned LC are flagged as potential flares. These undergo further validation criteria, e. g. that the decay time of the flare has to be longer than the rise time and that the flare maximum must not be the last flare point.   

For each validated flare, our algorithm determines the start time, peak time, flare duration, number of flare points, normalized flare amplitude, $A_\text{peak}$, and the equivalent duration, ED. The latter is obtained as the integral under the flare data points.

Bolometric flare luminosities, $L_\text{flare,bol}$, are calculated using Equations (1) and (5) of \cite{Shibayama2013}, and bolometric flare energies, $E_\text{flare,bol}$, from their Eq. (6). This approach assumes that the radiation in the {\it TESS} band of both the star and the flare can be modeled by a blackbody radiator. In the case of the star, the blackbody temperature is its effective temperature. In the case of the flare, common values reported in the literature range from $\approx8\,000$\,K to $14\,000$\,K (e. g. \citealt{Mochnacki1980}, \citealt{Kowalski2013}). We adopted a flare blackbody temperature of $T_\text{fl} = 10\,000$\,K. \cite{Shibayama2013} state that flare energies for $T_\text{fl}= 9\,000$\,K are about 66\% of those for $T_\text{fl} = 10\,000$\,K. This gives an estimate of the uncertainty.

\section{Analysis results}\label{sec:results}
\subsection{Flare incidence rate and activity level}\label{subsec:flare_incidence_activity}
\begin{figure}[t]
    \centering
    \includegraphics[width=0.48\textwidth]{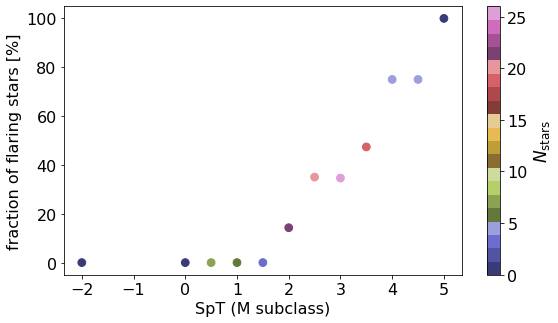}
    \caption{Flare incidence rate as a function of the SpT. Numbers on the x-axis denote M SpT subclasses. -2 stands for SpT K8.}
    \label{fig:fraction_flaring}
\end{figure}

We detected 2532 {\it TESS} flares on 35 stars of our sample, i. e. the overall flare incidence rate is $\approx32$\%\footnote{35/109 since 3 stars were excluded from the analysis.}. Fig.~\ref{fig:fraction_flaring} shows the flare incidence rate as a function of the SpT. It clearly increases from SpT M2 to M5 while the stars in our sample that have SpTs earlier than M2 did not show any flares at all.   

\begin{figure}[t]
    \centering
    \includegraphics[width=0.48\textwidth]{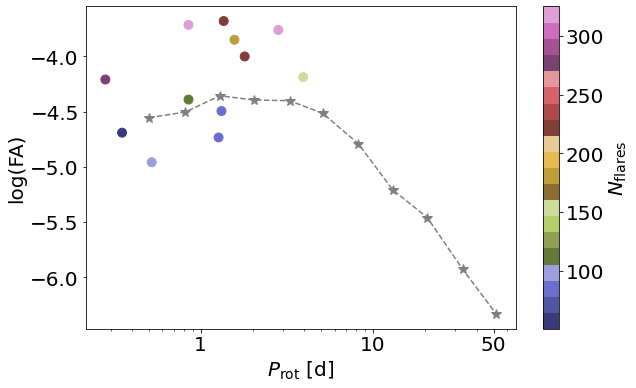}
    \caption{Flaring activity following \cite{Yang2017} as a function of rotation period. Overplotted in grey are the binned data points from Fig.~5 of \cite{Yang2017}.}
    \label{fig:flare_activity}
\end{figure}

Following \cite{Yang2017}, we define the `flare activity' as 
\begin{equation}
\text{FA}=(\sum_i E_{\text{flare,bol},i})/(L_\text{bol,*}\cdot t_\text{obs})
\label{eq:FA}
\end{equation}
This parameter represents the average flaring state of a star over the observed time interval. Here, $\sum E_{\text{flare,bol},i}$ is the sum of the bolometric energies of all flares detected for the respective target, $L_\text{bol,*}$ is its bolometric luminosity and $t_\text{obs}$ its total {\it TESS} observation time. Fig.~\ref{fig:flare_activity} shows the flare activity as a function of $P_\text{rot}$ for the 12 stars within our sample for which we detected a rotation period. Overplotted in grey are the binned values of \cite{Yang2017} (see their Fig.~5)\footnote{The data has been extracted with WebPlotDigitizer, https://automeris.io/WebPlotDigitizer/}. All our 12 stars are fast rotators with $P_\text{rot}<10$\,d, thus they lie in the saturated regime of the FA vs. $P_\text{rot}$ plot.

\subsection{Flare effects on potential planets and flare flux in the habitable zone}\label{subsec:flare_effects_HZ}
 As mentioned above, flares of appropriate energy and frequency can be beneficial for the development of life on exoplanets. Based on a laboratory study of \cite{Rimmer2018}, \cite{Guenther2020} derived an equation for the minimum flare frequency necessary to enable chemical reactions necessary to form the basic elements of prebiotic chemistry. This `abiogenesis zone' depends on the stellar radius and temperature and on the flare energy in the $U$ band (see Eq.~10 in \citealt{Guenther2020}). The equation is adapted to the case of a planet orbiting its host star at a distance at which the stellar flux is equal to the solar flux at Earth's orbit. 
 
 On the other hand, as a negative consequence, particularly highly energetic or frequent flare events can cause ozone depletion and, therefore, endanger life. \cite{Tilley2019} modeled the effects of repeated M dwarf flaring on an unmagnetized, Earth-like planet in the habitable zone. Assuming an orbital distance of 0.16\,AU -- which is the 1\,au equivalent orbital distance of AD Leo \citep{Segura2005} -- they modeled the flaring of a hypothetical host star using UV flare spectra of AD Leo obtained with the Hubble Space Telescope and flare energies and frequencies derived from {\it Kepler} data of GJ1243. Conservatively, they estimated the probability of an M dwarf flare to hit the orbiting planet to be 0.083. From this scenario, \cite{Guenther2020} derived a lower limit of $\nu>0.4$\,d$^{-1}$ on the frequency of flares with bolometric energies log$(E_\text{flare,bol})\text{ [erg]}>34$ to cause ozone depletion. A more permissive impact probability of 0.25 for each flare in the modeling of \cite{Tilley2019} results in $\nu>0.1$\,d$^{-1}$ according to \cite{Guenther2020}. 
 
\begin{figure}[t]
    \centering
    \includegraphics[width=0.48\textwidth]{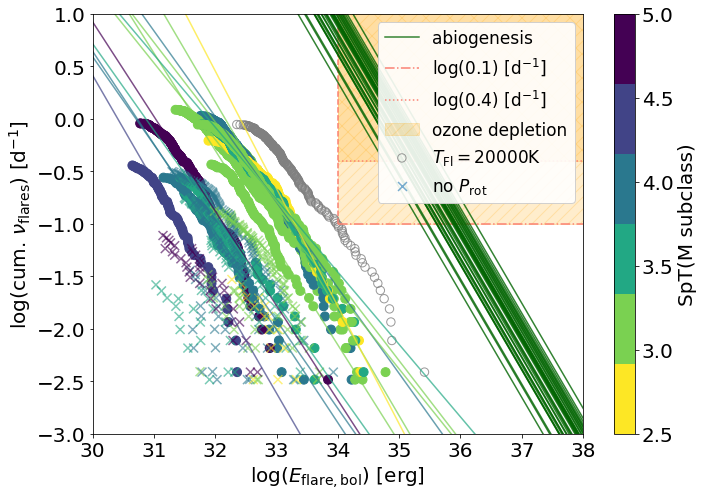}
    \caption{Cumulative flare energy frequency distributions for the 35 flaring stars in our sample. The green lines represent the abiogenesis zone of each star calculated following \cite{Guenther2020} (see Sect.~\ref{subsec:flare_effects_HZ} for details). In the orange area, highly energetic flares occur frequently enough to destroy a planet's ozone layer (see Sect.~\ref{subsec:flare_effects_HZ}). The grey `o'-markers denote the FFD for $T_\text{fl}=20\,000$\,K of TIC441734910 which is the target showing the highest flare energies in our sample. Even for this case, the area of ozone depletion is hardly entered.}
    \label{fig:FFDs}
\end{figure}

 To estimate the effect of flaring activity of our sample on potential planets in the HZ, Fig.~\ref{fig:FFDs} shows the cumulative flare energy frequency distributions (FFDs) in terms of bolometric flare energies for our sample. The power law fits are shown as solid lines in the same color as the respective FFD. Their slopes range from $-0.69$ to $-1.12$. For the fit, only data points above the energy completeness limit for flare detection are considered. This limit is determined following the approach of \cite{Hawley2014}. For details, see Bogner et al. (A\&A subm.). The lower limit of the abiogenesis zone for each star calculated following \cite{Guenther2020} is represented by a green line. The areas of ozone depletion for the permissive and the more conservative approach in \cite{Tilley2019} are colored in orange.         

\begin{figure}[t]
    \centering
    \includegraphics[width=0.48\textwidth]{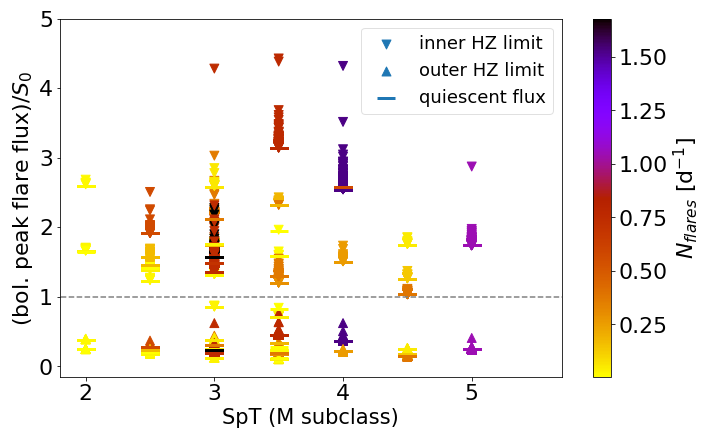}
    \caption{Bolometric peak flare flux at the inner and outer HZ limit as a function of SpT for the 35 flaring stars in our sample. The quiescent flux levels of each star at the inner and outer HZ boundaries are shown as a `\_'-marker. See Sect.~\ref{subsec:flare_effects_HZ} for details.}
    \label{fig:flare_flux_HZ}
\end{figure}

Since M dwarfs are cooler than the Sun, their HZs lie at orbital separations much smaller than 1\,au. This causes flares with energies comparable to those occurring on the Sun to have a greater impact on a planet in the HZ. Fig.~\ref{fig:flare_flux_HZ} shows for the 35 flaring stars in our sample the bolometric flux that reaches the inner and outer habitable zone boundary of the star at the peak of the flare. This flux is the sum of the bolometric stellar luminosity, $L_\text{bol,*}$, and the bolometric flare amplitude, $L_\text{flare,bol}$, divided by $4\pi d_\text{HZ}^2$, where $d_\text{HZ}$ is the orbital separation of the inner (Recent Venus, RV) or outer (Early Mars, EM) habitable zone limit. We calculated these HZ limits following \cite{Kaltenegger2019}. The flux in Fig.~\ref{fig:flare_flux_HZ} is normalized to the solar constant $S_0=1.361\cdot10^6$\,erg/(cm$^2$s). The quiescent flux level of each star is shown as a `\_'-marker in the respective color for both the RV and EM orbital separation. This shows that even without considering flares, the stellar flux at the inner HZ boundary is greater than $S_0$ for almost all stars. 
During flares, the flux at both HZ limits is additionally enhanced. This leads to a maximum total flux within our sample of $\approx4.5S_0$ at the inner HZ limit. The largest flux increase caused by a flare occurs on an M3 dwarf where the flare event raises the flux at the inner HZ boundary by $\approx3S_0$ (see Fig.~\ref{fig:flare_flux_HZ}). 

\subsection{Flare results of TOI and confirmed planet host stars}\label{subsec:flare_toi}
By matching our sample with the list of confirmed exoplanets and with the list of {\it TESS Objects of Interest} (TOI) from the \textit{NASA exoplanet archive}\footnote{exoplanetarchive.ipac.caltech.edu/ accessed 2021/06/19}, we found that 3 of our targets host a confirmed planet, namely TIC260004324 alias TOI-704 (\citealt{Gan2020}), TIC307210830 (L 98-59/TOI-175, \citealt{Kostov2019}, \citealt{2019A&A...629A.111C}) and TIC233193964 (GJ687, \citealt{Burt2014}). None of them shows rotational modulation, but for the latter two we detected flares. Four additional stars in our sample are classified as TOI: TIC198211976 (TOI-2283), TIC235678745 (TOI-2095), TIC260708537 (TOI-486) and TIC377293776 (TOI-1450). The latter two show flares.

\begin{figure}[t]
    \centering
    \includegraphics[width=0.48\textwidth]{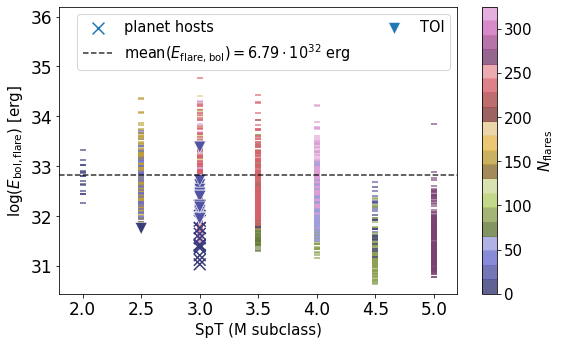}
    \caption{Bolometric flare energy as a function of the SpT. The number of validated flares for each star is color-coded. The flares of the 2 flaring planet host stars within our sample (both SpT M3) are shown as `x' markers, the 2 flaring TOI as `v' markers (see Sect.~\ref{subsec:flare_toi} for further information). The grey dashed line marks the average flare energy of the sample, $E_\text{bol,flare}\approx6.79\cdot10^{32}$\,erg.}
    \label{fig:E_flare_SpT}
\end{figure}

Fig.~\ref{fig:E_flare_SpT} shows the bolometric flare energy of all our 2532 validated flare events as a function of the SpT of the respective star. The TOI and confirmed planet hosts are highlighted as `v' and `x' symbols, respectively. In general, their bolometric flare energies are below the average of the sample. Only TIC377293776 exhibits one event above average with $E_\text{bol,flare}\approx2.44\cdot10^{33}$\,erg.   

\section{Summary and discussion}\label{sec:discussion}

We analyzed 1276 {\it TESS} primary mission 2-min. cadence light curves of 109 stars which were observed sufficiently long that the satellite can spot transiting planets in their entire habitable zone. The fraction of stars for which we found rotation periods is only 11\% which we ascribe to the high scatter in {\it TESS} light curves with respect to e. g. {\it K2} (see Bogner et al., A\&A subm.). All 12 stars for which we found a $P_\text{rot}$ are fast rotators with 0.28\,d $\leq P_\text{rot}\leq$ 3.93\,d. The overall flare incidence rate in our sample is $\approx32\%$. The fraction of flaring stars increases from SpT M2 to M5, while stars of SpTs earlier than M2 did not show any flares during the {\it TESS} observation.  

In absence of flare observations covering a broad wavelength range for our targets, we calculated bolometric flare energies and luminosities following the equations of \cite{Shibayama2013} that are based on a blackbody approximation for stellar and flare radiation. The fraction of the flare flux that falls into the {\it TESS} band is the same for all our flares since we assume a blackbody temperature of 10\,000\,K for all of them. The flux fraction covered {\it TESS} can be calculated via 

\begin{equation}
    \frac{\int R_\lambda B_\lambda(T_\text{fl})\text{d}\lambda}{\sigma_\text{SB}T_\text{fl}^4}\approx 17.6\%
\end{equation}

\noindent where $R_\lambda$ is the {\it TESS} response function\footnote{https://heasarc.gsfc.nasa.gov/docs/tess/data/tess-response-function-v1.0.csv}, $B_\lambda$ the planck function, $\sigma_\text{SB}$ the Stefan-Boltzmann constant and $T_\text{fl}=10\,000$\,K the blackbody temperature of the flare (see also \citealt{Schmitt2019}). 

We calculated the average flare activity during the {\it TESS} observation (see Eq.~\ref{eq:FA}), which had been introduced by \cite{Yang2017} as a new parameter quantifying the magnetic activity level of a star. From {\it Kepler} long cadence observations of 540 M dwarfs, \cite{Yang2017} showed that the FA follows the same qualitative behavior with the rotation period as other activity indicators (e. g. H$\upalpha$, X-ray luminosity), namely a division in a `saturated' regime of constant FA and a `correlated' regime in which the FA drops with increasing $P_\text{rot}$.
Since all stars in our sample are fast rotators, they lie in the saturated regime of the FA vs. $P_\text{rot}$ plot. Our FA values are in good agreement with those obtained by \cite{Yang2017} in the same $P_\text{rot}$ range. We thus confirm the typical FA level of log(FA)$\approx-4.5$ for M dwarfs in the saturated regime.

We calculated abiogenesis zones for each star following \cite{Guenther2020}. We then compared our bolometric FFDs to these abiogenesis zones as well as to the areas of ozone depletion given by \cite{Guenther2020} based on the irradiated planet atmosphere model of \cite{Tilley2019}. According to Fig.~\ref{fig:FFDs}, none of our stars exhibits sufficiently energetic and frequent flares to trigger ozone depletion or ribonucleotid synthesis. 

By calculating the bolometric peak flare flux in the HZ for each event, we found that the flux reaching the inner habitable zone limit (including the quiescent stellar flux) is up to $\approx4.5$ times as large as the quiescent solar flux reaching our Earth. The contribution of the largest flare increases the flux at the inner HZ boundary by $\approx3S_0$ as compared to the quiescent flux of the respective star. 

We note that modeling flare radiation as a 10\,000\,K blackbody is not accurate since the flare temperature is not constant throughout the event: \cite{Kowalski2013} found for simultaneous spectroscopic and photometric observations of 20 M dwarf flares in the optical and UV that during the rise phase, the blackbody temperature is increasing by $\sim2\,000$\,K. Also, e. g. \cite{Kowalski2019} pointed out that the optical emission from K-\&M-star flares deviates from a simple blackbody spectrum. According to these authors, a 9\,000\,K blackbody model underestimates the flare flux in the near-ultraviolet by a factor of 2. Additionally, as mentioned in Sect.~\ref{subsec:flare_detection}, the blackbody temperature of flares is not very well constrained. These caveats imply an uncertainty in our bolometric flare parameter calculations. 
To assess the effect of the unconstrained $T_\text{fl}$ on the bolometric FFDs, we determined them for several higher $T_\text{fl}$ values. We found that even assuming a blackbody temperature of $T_\text{fl}=20\,000$\,K for the flares, the FFD of the star showing the highest flare energies within our sample hardly enters the area of ozone depletion (see Fig.~\ref{fig:FFDs}).

Thus, judging by the bolometric flare energies calculated using the blackbody approach of \cite{Shibayama2013}, the flare events we recorded seem not to impact severely the planet's photochemistry. To further confirm this, simultaneous flare observations covering a broad wavelength range would be of use. Since no instruments exist yet to provide data for statistical flare studies in X-ray or UV wavelength on a grand scale, estimating X-ray flare energies on the basis of optical flare data as we did in Bogner et al. (A\&A submitted) is a promising approach to get further insight into the overall effect of flare events on potentially life-hosting planets.

\bibliography{references}%

\end{document}